# Performance Analysis of TCP Variants

**Dweep Gogia**
Khoury College of Computer and Information Sciences
Northeastern University
Boston, MA
**Email Id:** gogia.d@husky.neu.edu

**Mansukh Pamarath**
Khoury College of Computer and Information Sciences
Northeastern University
Boston, MA
**Email Id:** pamarath.m@husky.neu.edu

*Abstract –* The original design of the Transmission Control Protocol (TCP) worked reliably but was incapable of providing performance that could be scaled on bigger and larger congested networks. Several TCP variants have been proposed since then such as Reno, NewReno, Vegas, BIC, and so on, with features to control congestion. The objective of this project is to analyze these variants based on their performance, study and understand it's behavior under varying load and queueing algorithms.

## I. INTRODUCTION

There are various TCP variants such as Reno, Tahoe, Vegas, SACK and so on. These variants implement algorithms that handle congestion control. In our experiments we have used these variants to measure their performance such as throughput, delay (latency), and drop rate with respect to time and Constant Bit Rate (CBR) - no congestion control, specified bandwidth and sends packets at a specified rate. We have used the NS2 network simulator to perform all our experiments to analyze TCP performance. In the following sections we have elaborated the methodology, details, conditions and results of our experiments.

## II. METHODOLOGY

To begin, we have performed our experiments with the same network topology and configurations provided. The network topology used is shown in Fig. 3 Network configuration. In experiment 1 we have analyzed performance of TCP variants, using CBR rate. Hence, we vary the CBR rate until it reaches the bottleneck capacity of the network- that is when we begin to see drop in packets. Also, we keep parameters such as queue, delay, etc., unchanged. We use a tcl script to set up the network topology as shown in Fig. 3. Using this tcl file and passing parameters such as TCP variant and CBR rate in case of experiment 1 we generate trace files that contain data in the format shown in Fig. 1. Next, using a python script the trace file is parsed where all the calculations for various conditions such as throughput, delay, drop rate are calculated from the trace file data. Finally, we generate a graph using gnuplot for the experiments we have performed. Throughout this paper you will find graphical representations of the results of our experiments.

## III. TOOLS USED

For carrying out the three experiments, we have used NS-2 simulator to simulate our network configuration. Each NS-2 simulation is controlled by a script file written in TCL. NS-2 is a useful tool for representing the network traffic graphically and supports several queueing and routing algorithms. Further, NS-2 comes with support for multiple protocols for generating network flow [3]. We have generated the trace file used for analyzing our network by writing TCL scripts.

### A. Generating Trace File

Following are the steps involved in generating a trace file:
- Declare Simulator and setting output file
- Setting Node and Link (network configuration)
- Setting Agent
- Setting Application
- Setting Simulation time and schedules
- Declare finish

### B. Analyzing Trace File

A sample part of a trace file is shown in figure 1. Each line in the trace file gives information about the packet. Description of what each column represents is given in figure 2.

```
r 1.3556  3 2 ack  40 ------- 1 3.0 0.0 15 201
+ 1.3556  2 0 ack  40 ------- 1 3.0 0.0 15 201
- 1.3556  2 0 ack  40 ------- 1 3.0 0.0 15 201
r 1.35576 0 2 tcp 1000 ------- 1 0.0 3.0 29 199
+ 1.35576 2 3 tcp 1000 ------- 1 0.0 3.0 29 199
d 1.35576 2 3 tcp 1000 ------- 1 0.0 3.0 29 199
+ 1.356   1 2 cbr 1000 ------- 2 1.0 3.1 157 207
- 1.356   1 2 cbr 1000 ------- 2 1.0 3.1 157 207
```

Fig. 1.  Sample part of trace file

| event | time | from node | to node | pkt type | pkt size | flags | fid | src addr | dst addr | seq num | pkt id |

Fig. 2.  Column values of the trace file

The first column is the event type of packet. Following are the possible events [4]:

- + : Packet enters to a queue
- - : Packet exits a queue
- r : Packet is received by a node
- d : Packet is dropped

To parse the data generated we have used Python scripts. Python is a common interpreted programming language. Gnuplot is a portable command line utility, it supports various 2D and 3D plots over wide range of platforms.

## IV. NETWORK CONFIGURATION

The network topology that we have used for all the three experiments is shown in figure 3. Each of the links has a bandwidth of 10Mbps for all three experiments. We alter the protocols, CBR rate and use various queueing algorithms to perform the experiments using the given network topology.

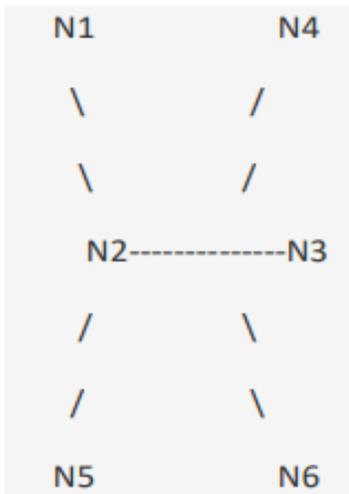

Fig. 3. Network configuration

### A. Experiment 1 - TCP Performance Under Congestion

For this experiment, we have a single TCP flow from N1 to a sink at N4. Additionally, there is a CBR source at N2 and a sink at N3. We use an FTP application on TCP to generate traffic.

### B. Experiment 2 - Fairness Between TCP Variants

We have three flows in this experiment: one CBR, and two TCP. A CBR source at N2 and a sink at N3, and two TCP streams from N1 to N4 and N5 to N6, respectively. Again, we use an FTP application on both TCP flows to generate traffic.

### C. Experiment 3 - Influence of Queuing

Using the same network topology, we have one TCP flow with a source at N1 and sink at N4 and one CBR/UDP flow from N5 to N6. We also set up an FTP traffic generator on the TCP flow.

## V. RESULT OF EXPERIMENT 1

In the first set of experiments, we analyze the performance of various TCP variants (Tahoe, Reno, NewReno, and Vegas) under the influence of various load conditions. We use the network topology given in figure 3, with the bandwidth of each link being constant at 10Mbps. We analyze the performance of the TCP variants listed above in the presence of a Constant Bit Rate (CBR) flow. The network configuration for these experiments is described in section 4-A.

We analyze the throughput, packet drop rate, and latency of the TCP stream as a function of the bandwidth used by the CBR flow. That means we are varying the bandwidth of the CBR flow to record the performance of various TCP flows under congestion conditions. We are changing the CBR bandwidth from 1Mbps to 12Mbps and recording the performance of TCP variants by plotting a graph. For each graph, the X-axis represents CBR rate and Y-axis represents either of throughput, packet drop rate or latency.

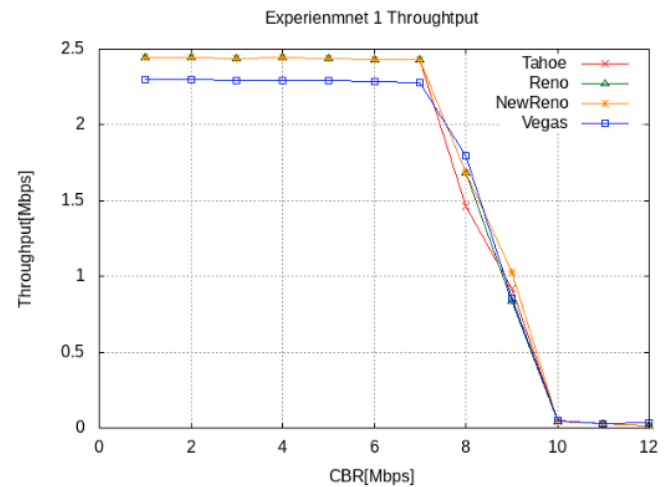

Fig. 4. Experiment 1 - Throughput

The graph in figure 4 shows the throughput of Tahoe, Reno, NewReno, and Vegas under CBR rates varying from 1Mbps to 12Mbps. Initially, until about CBR of 7Mbps, the throughput of all the TCP variants is almost the same and is constant. This point can be considered as the cliff of the flow as after 7Mbps the throughput of all variants tends to 0. Also, one can see that initially, Vegas started off with lower throughput as compared to other variants. But as CBR rate is increased, after the cliff point at CBR 12Mbps, Vegas recovers better from the congestion than the other variants.

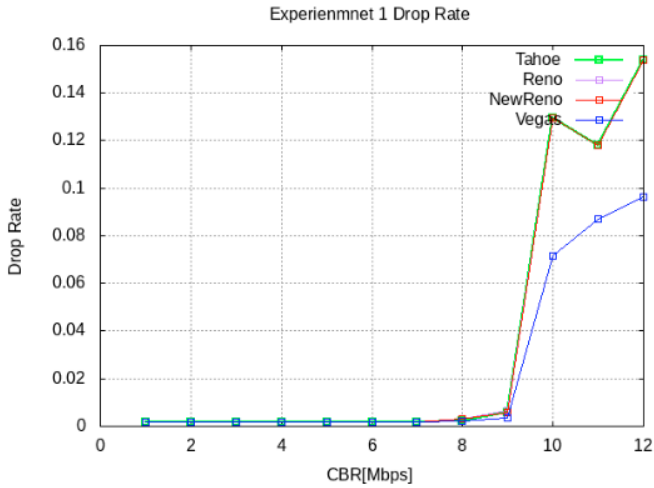

Fig. 5. Experiment 1 – Drop Rate

The graph in figure 5 shows the drop rate of TCP variants under varying CBR flow. The drop rate for all the variants is constant and low until 9Mbps CBR rate. After that, it increases exponentially for all the variants. However, after 10Mbps rate, we can see that the drop rate for Tahoe, Reno, and NewReno continues to increase exponentially but the rate of increase of drop rate for Vegas starts decreasing. We believe this is because of the difference in the TCP Congestion Avoidance algorithm used by Vegas than the other three variants [5]. Vegas uses packet delay, rather than packet loss and hence, is able to control congestion earlier and perform actions for avoiding congestion.

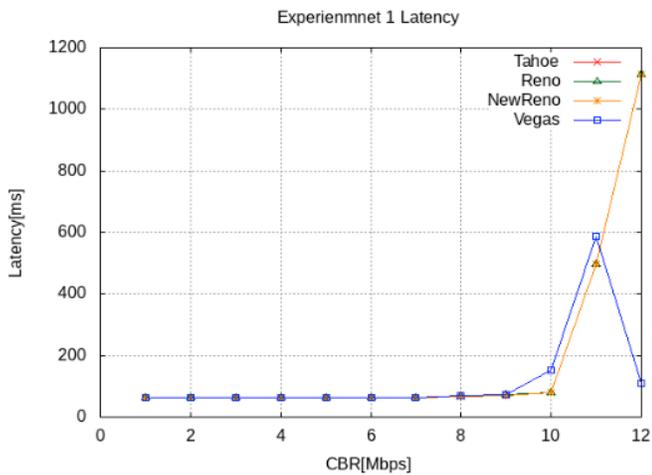

Fig. 6. Experiment 1 – Latency

Figure 6 shows the latency of TCP variants under varying CBR flow. Similar to the throughput and drop rate, latency is constant and low for all variants until a certain CBR flow (10Mbps). As the CBR rate is increased, the latency for all variants increases almost exponentially. However, because of the better algorithm that Vegas uses for congestion avoidance, as discussed above in drop rate, it is able to improve its latency quite considerably after 11Mbps, while for other it continues to increase [6].

We can, therefore, conclude that as far as experiment 1 is concerned, Vegas is the best TCP variant amongst the other three. This is because of the efficient congestion avoidance algorithm Vegas uses to detect and act against congestion. We will now see the performance of all the TCP variants under more than one flows and queueing algorithms in the following sections.

## VI. RESULT OF EXPERIMENT 2

In the second experiment, we have analyzed the fairness of pairs of TCP variants. The configurations for this experiment we made – CBR source at N2, sink at N3, two TCP streams that start at N1 to N4 and N5 to N6. Here, again we calculate the throughput, latency and drop rate between the two pairs of TCP variants against CBR flow rate. We performed this experiment on these pairs of TCP variants [1]:

1. Reno and Reno
2. NewReno and Reno
3. Vegas and Vegas
4. NewReno and Vegas

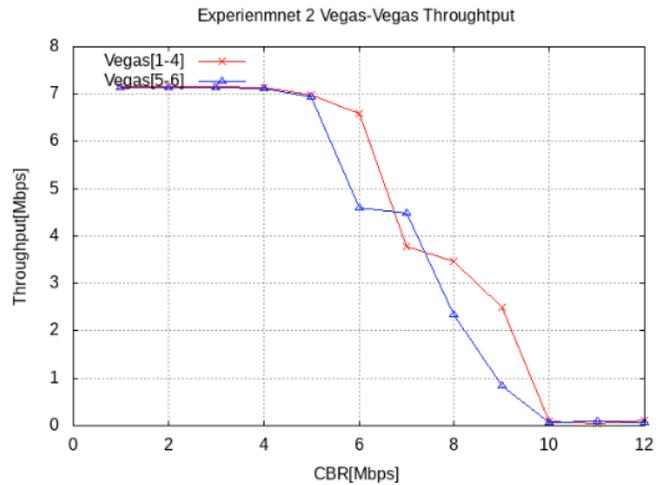

Fig. 7. Experiment 2 – Throughput of Vegas and Vegas

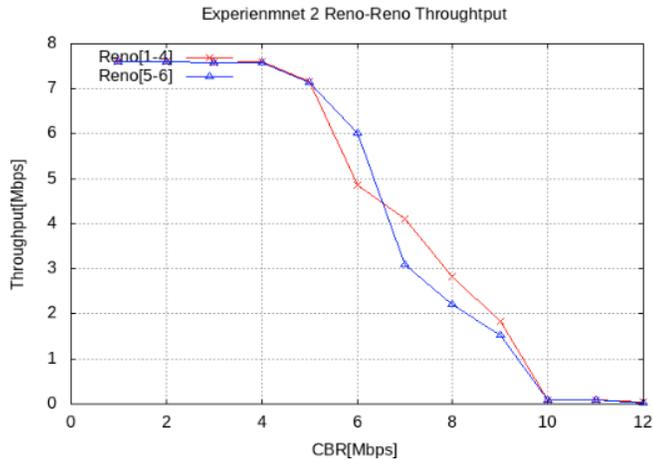

Fig. 8. Experiment 2 – Throughput of Reno and Reno

The above figures (Fig. 7. and Fig. 8.) are the results of our experiment 2. The throughputs of Vegas/Vegas and Reno/Reno are similar to each other. Whereas, the combination of different variants are unfair to each other. However, when we compare the results of NewRen/Reno throughput, we find that the throughput of NewReno performs better compared to that of Reno. We calculated the average throughputs for both the case scenarios and found out that the difference for the Reno/Reno variant was just around 0.03 Mbps. Similarly, the average packets dropped in case of Reno/Reno was about 0.2. In contrary in the case of NewReno and Reno, NewReno get more bandwidth compared to Reno hence, it performs better. When we look into the implementations of these TCP variants, NewReno and Reno differ in terms of the fast-recovery. NewReno does'nt exit fast recovery until all the data which was out standing at that time when it entered fast-recovery received all the acknowledgements. Reno reduces the congestion window multiple times, which is corrected in NewReno [2]. Also, we notice that in NewReno and Reno, the congestion is detected only after the packets start dropping. Whereas, in Vegas, it detects it in the initial stage itself.

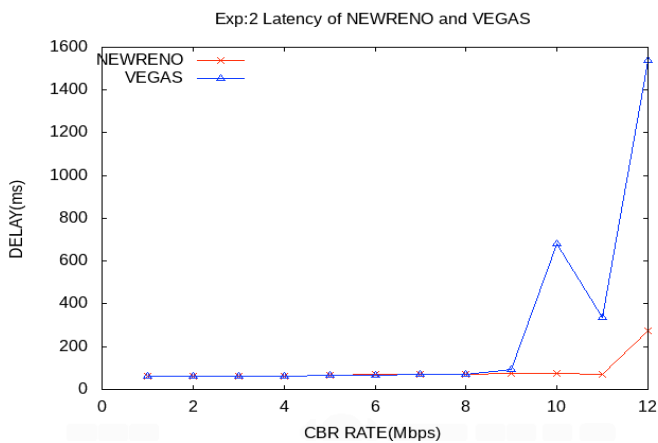

Fig. 9. Experiment 2 – Latency of NewReno and Vegas

To summarise, the performance of NewReno is better than Reno as it doesn't exit fast-recovery phase like in Reno, hence this gives NewReno greater bandwidth. In the case of NewReno and Vegas, the latter reduces the transmission rate to avoid congestion before it occurs. Whereas, the former keeps sending packets until congestion occurs.

## VII. RESULT OF EXPERIMENT 3

In the third set of experiments, instead of varying the rate of the CBR flow, we will study the influence of the queuing discipline used by nodes on the overall throughput of flows. We are using two queueing algorithms which are DropTail and Random Early Drop (RED) that control how packets in a queue are treated. We use the network topology given in figure 3, with the bandwidth of each link being constant at 10Mbps. The network configuration for these experiments is described in section 4-C.

We first start the TCP flow. Once the TCP flow is steady, we start the CBR source and analyze how the TCP and CBR flow changes under the following queuing algorithms: DropTail and RED. We are using TCP Reno and SACK for this set of experiments. Unlike experiments 1 and 2, here we are not varying the CBR flow, instead, we are analyzing the performance of TCP and CBR flow over time. Hence, for each graph, X-axis represents time in seconds and Y-axis represents throughput or latency.

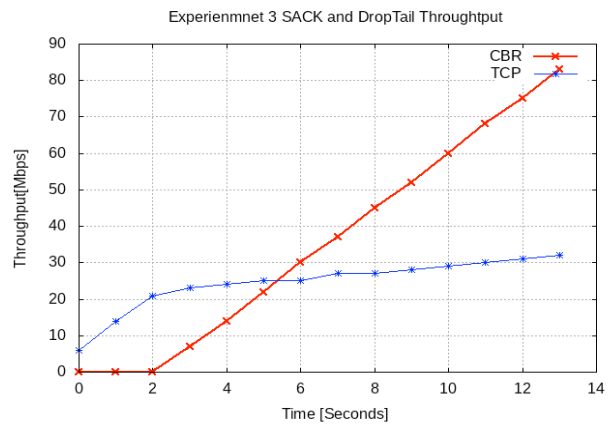

Fig. 10. Experiment 3 – SACK and DropTail Throughput

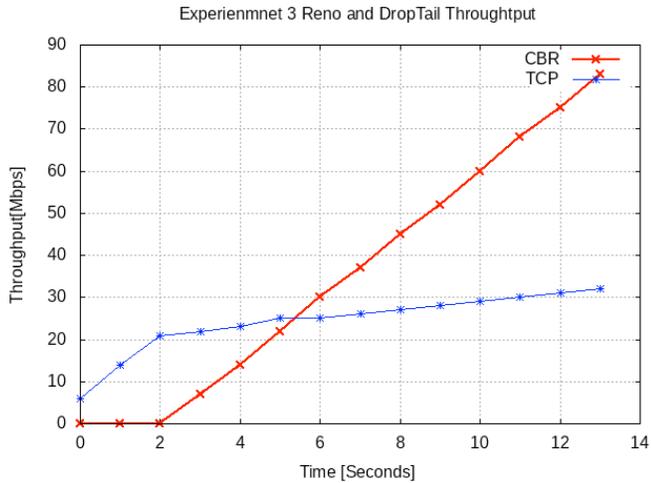

Fig. 11.  Experiment 3 – Reno and DropTail Throughput

With figure 7 and 8 we can conclude that the DropTail algorithm does not provide fair bandwidth to each flow. The TCP flow in both graphs tends to decline after the creation of the CBR flow. Also, the latency of both TCP variants increases abruptly with the DropTail algorithm as shown in figure 9 and 10. On average, the latency of both TCP flows is higher with DropTail algorithm as opposed to RED. As time increases, latency with the RED algorithm tends to remain low constantly which is not the case with the DropTail algorithm.

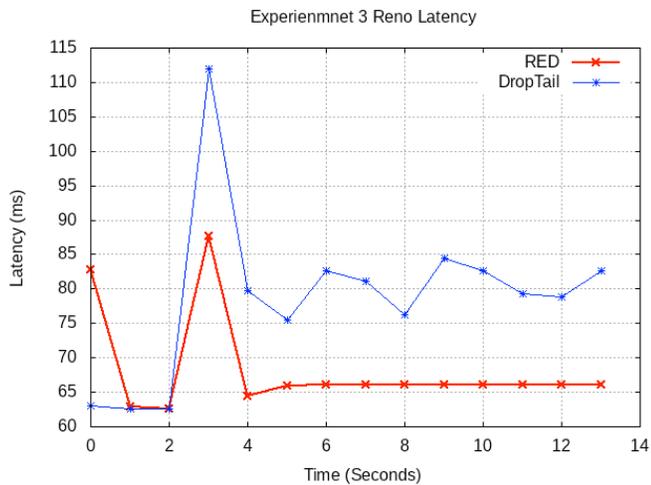

Fig. 12.  Experiment 3 – Reno Latency

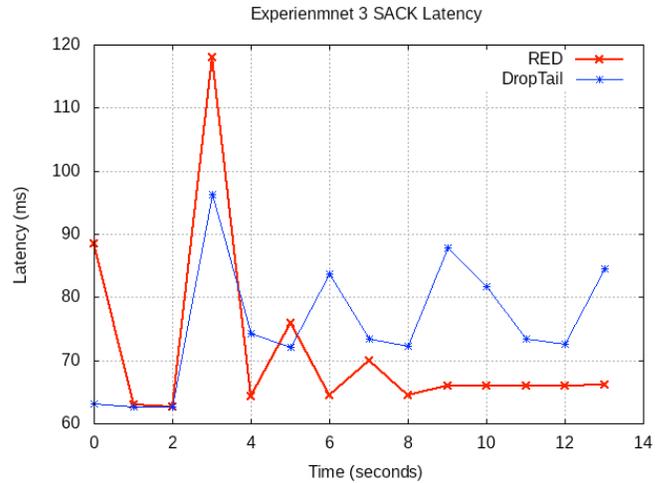

Fig. 13.  Experiment 3 – SACK Latency

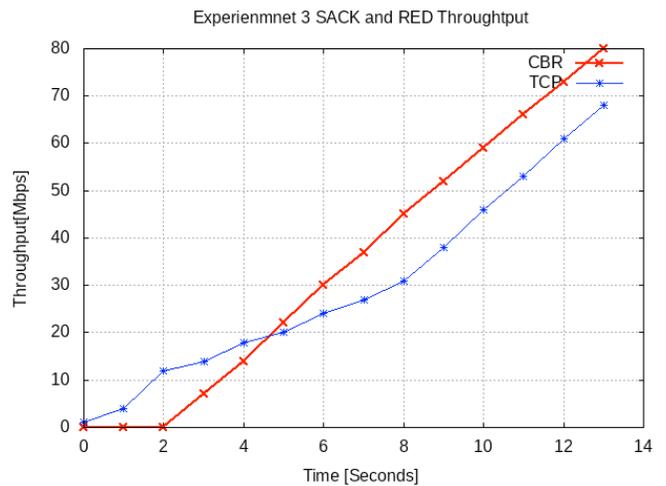

Fig. 11.  Experiment 3 – SACK RED Throughput

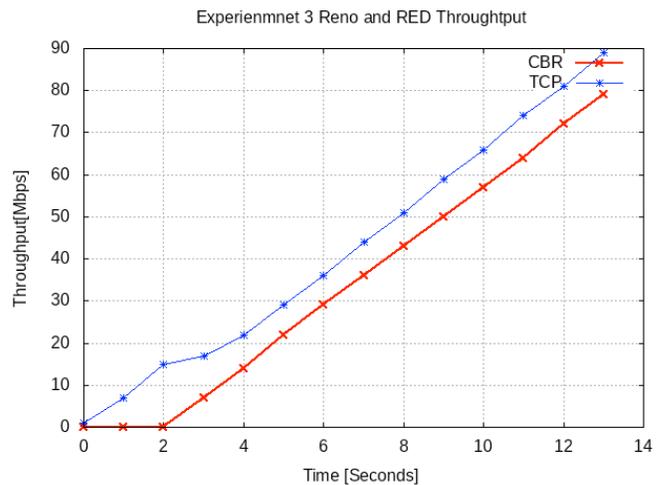

Fig. 14.  Experiment 3 – Reno RED Throughput

Graphs in figure 11 and 12 show that the RED algorithm provides a better and fairer bandwidth to both the flows. The TCP flow increases steadily even after the creation of CBR flow.

Therefore, we can safely conclude that RED provides fairer bandwidth to each flow as compare to DropTail. RED also provides lower latency for both TCP variants, on average. Further, with figure 10 and 11, we can show that RED is a favorable algorithm when dealing with TCP SACK.

## VIII. CONCLUSION

With the set of experiments listed above, we performed a complete analysis of TCP variants under various load conditions and queueing algorithms. We conclude the following from the three experiments:
- Experiment 1 - TCP Performance Under Congestion: Amongst the four TCP variants, Vegas is the stand out as far as experiment 1 is concerned. It has a higher throughput, lower latency and drop rate as compared to other variants. This is because of the efficient congestion avoidance algorithm that Vegas uses.
- Experiment 2 - The performance of variants when compared against each other are very similar and hence are fair to each other. Whereas from the graphs and explanations provided in the experiment 2, we can conclude that the performance of different variants with each other are unfair. NewReno and Vegas proves a good example in this case
- Experiment 3 - Influence of Queuing: With the third set of experiments, we know that queueing algorithms like RED and DropTail do affect the performance of network flows including TCP and CBR. We also see that RED queueing algorithm provides fairer bandwidth and lower latency as compared to the DropTail algorithm Thus, RED is a good choice when dealing with TCP SACK.